\documentclass[useAMS,usenatbib]{mn2e}
\usepackage{graphicx}
\title{Identifying dark matter signals by the radio continuum spectral data of the cool-core cluster RX J1720.1+2638}
\author[Chan \& Lee]{Man Ho Chan \thanks{chanmh@eduhk.hk}, Chak Man Lee
\\ Department of Science and Environmental Studies, The Education University of Hong Kong, Tai Po, Hong Kong}

\begin{document}

\date{Accepted XXXX, Received XXXX}

\pagerange{\pageref{firstpage}--\pageref{lastpage}} \pubyear{XXXX}

\maketitle

\label{firstpage}

\date{\today}

\begin{abstract}
Investigating the signals of dark matter annihilation is one of the most popular ways to understand the nature of dark matter. In particular, many recent studies are focussing on using radio data to examine the possible signals of dark matter revealed in galaxies and galaxy clusters. In this article, we investigate on the spectral data of the central radio halo of the cool-core cluster RX J1720.1+2638. We show that the radio spectral data can be best accounted by the synchrotron emission due to dark matter annihilation via $\tau$ lepton channel (with dark matter mass $m=15$ GeV) or $b$ quark channel (with dark matter mass $m=110$ GeV), although using the very coarse spectral data with notable errors. Despite the fact that cosmic-ray emission can also provide a good explanation for the observed radio spectrum, our results suggest a possible positive evidence for dark matter annihilation revealed in the form of radio emission in RX J1720.1+2638 cluster.
\end{abstract}

\section{Introduction}
The nature of dark matter remains a mystery in particle astrophysics and cosmology. It is commonly believed that dark matter consists of unknown massive particles which almost do not interact with ordinary matter. A kind of particles called Weakly Interacting Massive Particles (WIMPs) are currently one of the most popular candidates to account for the dark matter problem \citep{Feng,Arcadi}. Many new studies have been initiated to search for the signals of WIMPs \citep{Bringmann,Fukuda}.

Theories predict that WIMPs might self-annihilate to give high-energy stable particles such as electrons, positrons, neutrinos, and photons \citep{Arcadi}. For different annihilation channels, the energy spectrum of these high-energy stable particles can be uniquely determined by dark matter mass \citep{Cirelli}. Therefore, by detecting the energy spectrum of the high-energy stable particles produced by dark matter annihilation, one can probably identify the indirect signal of dark matter and probe the properties of dark matter.

Some traditional measurements have been done for detecting the indirect signal of dark matter, such as gamma-ray detection \citep{Abdo,Albert} and cosmic-ray detection \citep{Aguilar,Ambrosi}. Although some studies have claimed the existence of a gamma-ray excess from our Galactic centre \citep{Calore,Abazajian,Daylan}, whether these excess gamma rays purely originate from dark matter annihilation is still controversial because the actual gamma-ray contribution of pulsars at the Galactic centre is poorly known \citep{Buschmann,Ye}. Apart from gamma-ray and cosmic-ray detection, radio analysis is another way to study the signal of dark matter annihilation. Many previous radio studies focussing on different structures, such as small galaxies \citep{Kar,Regis,Chan}, nearby large galaxies \citep{Egorov,Chan2,Chan3}, and galaxy clusters \citep{Storm,Chan4,Chan5,Beck,Lavis}, have put efforts on constraining dark matter and identifying dark matter signals. Generally speaking, radio analysis using different observing frequencies is able to identify the possible indirect signal of dark matter. It is because the energy spectrum of the high-energy stable electrons or positrons produced by dark matter annihilation is highly correlated with the radio spectrum. Therefore, by analysing the multi-frequency radio spectrum of a structure, one can possibly determine the signal of dark matter by investigating on the shape of the radio continuum spectrum \citep{Chan4,Chan}.

In this article, we perform a multi-frequency radio analysis of dark matter using the data of the distant galaxy cluster RX J1720.1+2638. This galaxy cluster has a small radio halo at its centre with radius $r_{\rm h} \approx 70$ kpc \citep{Giacintucci,Biava} and it also shows significant extended emission on much larger scales up to about 600 kpc \citep{Savini,Biava,Biava2}. These features might possess the information of dark matter annihilation. Subject to the systematic uncertainties in the cosmic-ray emission modelling, we show that the radio spectrum of the central small radio halo can be best accounted by two possible dark matter annihilation scenarios: 1. dark matter mass $m=15$ GeV via the $\tau$ lepton channel, or 2. dark matter mass $m=110$ GeV via the $b$ quark channel. Therefore, our results provide slightly positive evidence showing the signal of dark matter annihilation in RX J1720.1+2638 cluster.

\section{Radio emission model of dark matter annihilation}
The high-energy electrons and positrons produced from dark matter annihilation would interact with magnetic field to give synchrotron radiation. For WIMP mass with the order of $m \sim 10-100$ GeV, the energy of the electrons or positrons would be about $\sim 1$ GeV. This would give the synchrotron radiation in mainly radio bands with frequency $\nu \sim 1$ GHz. Therefore, we can theoretically transform the energy spectrum of the high-energy particles to the predicted radio spectrum. In the following, we will describe the radio emission model of dark matter annihilation for the galaxy cluster RX J1720.1+2638.

The average power at frequency $\nu$ under magnetic field $B$ for synchrotron emission induced by the dark matter annihilation is given by \citep{Storm}
\begin{equation}
P_{\rm syn}(\nu)=\int_0^{\pi}d\theta\frac{(\sin\theta)^2}{2}2\pi\sqrt{3}r_{\rm e}m_{\rm e}c\nu_{\rm g}F_{\rm syn}\left(\frac{x}{\sin\theta}\right),
\end{equation}
where $\nu_{\rm g}=eB/(2\pi m_{\rm e}c)$, $r_{\rm e}$ is the classical electron
radius, and $F_{\rm syn}(x/\sin\theta)=x/\sin\theta\int_{x/\sin\theta}^{\infty}K_{5/3}(s)ds$ with the quantity $x$ defined as
\begin{equation}
x=\frac{2\nu(1+z)}{3\nu_{\rm g}\gamma^2}\left[1+\left(\frac{\gamma\nu_{\rm p}}{\nu(1+z)}\right)^2\right]^{3/2},
\end{equation}
in which $z$ is the redshift, $\gamma$ is the Lorentz factor of the electrons or positrons, and $\nu_{\rm p}=8890[n(r)/1\;{\rm cm}^{-3}]^{1/2}$ Hz is the plasma frequency with the number density of the thermal electrons $n(r)$.

Apart from synchrotron cooling, the high-energy electrons and positrons would also cool down via inverse Compton scattering, bremsstrahlung and Coulomb losses. The total cooling rate $b_T(E,r)$ contributed by these four processes can be explicitly expressed (in unit of $10^{-16}$ GeV s$^{-1}$) as \citep{Egorov2}
\begin{eqnarray}
b_T(E,r)&=&0.0254E^2B^2 \nonumber\\
&&+ 0.25E^2(1+z)^4 \nonumber\\
&&+ 7.1\times10^{-4}\gamma n(r)\left(0.36+\ln\gamma\right)\nonumber\\
&&+7.6\times10^{-2} n(r)\left[73+\ln\left(\frac{\gamma}{n(r)}\right)\right].
\label{cooling}
\end{eqnarray}
Here, the magnetic field strength $B$ and the energy of electrons and positrons $E$ are in $\mu$G and GeV respectively. For an electron with energy $E=1$ GeV travelling by $d=10$ kpc in a typical galaxy cluster, the diffusion time would be $t_d \sim d^2/D \sim 1\times 10^{17}$ s, where we have assumed a typical diffusion coefficient $D \sim 10^{28}$ cm$^2$/s. Nevertheless, the cooling time for a 1 GeV electron would be $t_c \sim E/b_T \sim 3 \times 10^{15}$ s, with a typical value $B\sim 10$ $\mu$G. Since the diffusion time scale inside a galaxy cluster is much longer than the cooling time scale of the high-energy electrons and positrons, the diffusion process is insignificant in the radio emission process. In this situation, the high-energy electron and positron energy spectrum distribution can be well-approximated by the following equilibrium state \citep{Storm}:
\begin{equation}
\frac{dn_e}{dE}= \frac{\langle \sigma v \rangle[\rho_{\rm DM}(r)]^2}{2m^2b_T(E,r)} \int_E^m \frac{dN_{\rm e,inj}}{dE'}dE',
\end{equation}
where $\langle \sigma v \rangle$ is the annihilation cross section, $\rho_{\rm DM}(r)$ is the dark matter density profile, and $dN_{\rm e,inj}/dE'$ is the injection energy spectrum of electrons and positrons due to dark matter annihilation. The injection energy spectrum for different annihilation channels can be found in \citet{Cirelli}.

The X-ray surface brightness profile of many galaxy clusters can be best-described by the so-called single-$\beta$ model \citep{Cavaliere,Cavaliere2,Ettori,Chen}. The thermal electron number density distribution can be modelled by the following profile \citep{Chen}:
\begin{equation}
n(r)=n_0\left(1+\frac{r^2}{r_{\rm c}^2}\right)^{-3\beta/2},
\end{equation}
where $n_0$, $r_{\rm c}$  and $\beta$ are empirical parameters in the fitting process. In Fig.~1, we show the updated data of the thermal electron number density profile (observed from the {\it Chandra} X-ray Observatory) for RX J1720.1+2638 presented in \citet{Cavagnolo}, which can be best fitted with the single-$\beta$ model using $n_0=0.08$ cm$^{-3}$, $r_{\rm c}=21.7$ kpc, and $\beta=0.46$. We have also fitted the electron number density profile by the double-$\beta$ model \citep{Chen}. Nevertheless, the double-$\beta$ model does not give a significant improvement in the fitting (less than 1\% improvement in the $\chi^2$ value), though this model involves three more free parameters. As our whole analysis is not sensitive to the electron number density profile, we simply apply the single-$\beta$ model to minimise the number of free parameters. For the magnetic field profile, as shown by previous studies that it can be traced by the thermal electron number density, we write it in the following form \citep{Storm,Govoni}:
\begin{equation}
B(r)=B_0\left[\left(1+\frac{r^2}{r_{\rm c}^2}\right)^{-3\beta/2}\right]^{\eta},
\end{equation}
where $\eta=0.5-1.0$ and $B_0$ is the central magnetic field which can be determined by \citep{Kunz,Govoni}
\begin{equation}
B_0=11\epsilon^{-0.5}\left(\frac{n_0}{0.1\,{\rm cm}^{-3}}\right)^{0.5}\left(\frac{T_0}{2\,{\rm keV}}\right)^{3/4}\,\mu {\rm G}.
\label{Boestimate}
\end{equation}
Here, $T_0$ is the central hot gas temperature and $\epsilon=0.5-1.0$ is the index modelled in simulations. Therefore, at the central region of RX J1720.1+2638 with $T_0=4.08$ keV and $n_0=0.08$ cm$^{-3}$ \citep{Cavagnolo,Jimenez}, we get $B_0=16.2-24.6$ $\mu$G for the possible range of $\epsilon=0.5-1.0$. This range of central magnetic field is reasonable as there are some galaxy clusters (e.g. A2029, A2052, A2199, 3C129.1, Cygnus A) possibly having $B_0$ larger than 10 $\mu$G \citep{Eilek,Kunz}. A few galaxy clusters such as Virgo, A4059, and A780 might have $B_0$ larger than 30 $\mu$G \citep{Eilek}. Nevertheless, the value of $B_0$ here is somewhat larger than the values reported in some nearby galaxy clusters, such as Coma cluster (4-5 $\mu$G) \citep{Bonafede} and Fornax cluster ($<10$ $\mu$G) \citep{Anderson}. The larger value of $B_0$ calculated in RX J1720.1+2638 is due to its relatively higher central thermal electron number density $n_0$. The relation in Eq.~(7) has been shown to give good agreements with observations \citep{Kunz}. Note that the average magnetic field strength for RX J1720.1+2638 could be smaller than 10 $\mu$G as $B_0$ represents the central magnetic field strength. 

RX J1720.1+2638 cluster is a cool-core galaxy cluster, in which the hot gas temperature is not uniform throughout the cluster. Based on the consistent observational data obtained from {\it XMM-Newton} and {\it Chandra} \citep{Jimenez,Cavagnolo,Biava2}, the hot gas temperature of RX J1720.1+2638 can be parameterized as follow \citep{Allen}:
\begin{equation}
T(r)=T_0+T_1\left[\frac{1}{1+(r/r_0)^{-a}}\right],
\end{equation}
where the best-fit parameters are $T_0=4.0$ keV, $T_1=3.9$ keV, $r_0=134$ kpc, and $a=1.76$ (see Fig.~2). Using the temperature profile of RX J1720.1+2638, we can get the final important ingredient of the model - the dark matter density distribution. Theoretically, if the hot gas is in hydrostatic equilibrium, one can obtain the dark matter distribution directly by putting the thermal electron density profile $n(r)$ and the temperature profile $T(r)$ into the hydrostatic equation to get:
\begin{equation}
\rho_{DM}(r) \approx \frac{1}{4\pi r^2}\frac{d}{dr}\left[-\frac{kT(r)r}{\mu m_{\rm p}G}\left(\frac{d\ln n(r)}{d\ln r}+\frac{d\ln T(r)}{d\ln r}\right)\right],
\label{lndensity}
\end{equation}
where $\mu=0.59$ is the molecular weight and $m_{\rm p}$ is the proton mass. Here, we have assumed that dark matter has dominated the density distribution in RX J1720.1+2638. However, RX J1720.1+2638 is a cool-core cluster in which the systematic uncertainties of using hydrostatic equilibrium is not negligible near the centre of the galaxy cluster \citep{Biffi}. It is clear from X-ray observations that cold fronts and a sloshing spiral exist within the intracluster medium \citep{Mazzotta,Biava2}. This shows that the cluster deviates from hydrostatic equilibrium. Also, the hydrostatic mass profile in Eq.~(9) sensitively depends on the gradient of the central temperature profile, which contains a relatively large uncertainty for most cool-core clusters \citep{Vikhlinin,Hudson}. In fact, the hydrostatic mass profile can be a very good approximation to the dynamical mass profile outside the core region of a galaxy cluster. However, for the core region, using Eq.~(9) would give a relatively large deviation. Fortunately, numerical simulations \citep{Navarro} and observations \citep{Pointecouteau} suggest a kind of universal density profile called the Navaro-Frenk-White (NFW) profile which can generally model the dark matter density profile of a galaxy cluster. The universal NFW dark matter density profile is given by \citep{Navarro}:
\begin{equation}
\rho_{\rm DM}(r)=\frac{\rho_{\rm s}r_{\rm s}^3}{r(r+r_{\rm s})^2},
\end{equation}
where $\rho_{\rm s}$ and $r_{\rm s}$ are the scale density and scale radius respectively. As we need to focus on the core region of RX J1720.1+2638, we adopt the NFW profile to describe the dark matter distribution. We use the NFW profile to fit with the hydrostatic profile in Eq.~(9) outside the core region of RX J1720.1+2638 ($r \ge 100$ kpc) to get the best-fit NFW scale density $\rho_{\rm s}=8.26\times 10^5M_{\odot}$ kpc$^{-3}$ and scale radius $r_{\rm s}=423$ kpc. In Fig.~3, we can see that the NFW profile agrees with the hydrostatic profile very well outside the core region. Significant deviation between the NFW profile and hydrostatic profile can be found in $r<70$ kpc, which has almost covered the region of the central radio halo. Moreover, we also involve two other popular dark matter density profiles, the Moore profile \citep{Moore} and Einasto profile \citep{Einasto,Springel}, in our analysis. These profiles are respectively given by
\begin{equation}
\rho_{\rm DM}(r)=\rho_{\rm M}\left[\left(\frac{r}{r_{\rm M}}\right)^{1.5}\left(1+\left(\frac{r}{r_{\rm M}}\right)^{1.5}\right)\right]^{-1}
\end{equation}
where $\rho_{\rm M}=9.93 \times 10^4 M_{\odot}$ kpc$^{-3}$ and $r_{\rm M}=851$ kpc, and
\begin{equation}
\rho_{\rm DM}(r)=\rho_{\rm e}\exp\left\{-\frac{2}{\alpha_{\rm e}}\left[\left(\frac{r}{r_{\rm e}}\right)^{\alpha_{\rm e}}-1\right]\right\}
\end{equation}
where $\alpha_{\rm e}=0.3354$, $\rho_{\rm e}=2.77 \times 10^5 M_{\odot}$ kpc$^{-3}$, $r_{\rm e}=363$ kpc. These profiles can also match with the hydrostatic profile outside the core region (see Fig.~3).

Assume that the dark matter distribution is spherically symmetric, and the luminosity distance to RX J1720.1+2638 cluster $D_{\rm L}=765.4$ Mpc \citep{Giacintucci} is sufficiently large, the radio flux density contributed by dark matter annihilation is finally given by
\begin{equation}
S_{\rm DM}(\nu)=\frac{2(1+z)}{4\pi D_{\rm L}^2} \int_0^{r_{\rm h}}\int_{m_{\rm e}}^{m}\frac{dn_{\rm e}}{dE}P_{\rm syn}(\nu)dE(4\pi r^2)dr.
\label{SDM}
\end{equation}
The factor two in Eq.~(\ref{SDM}) indicates the contribution of both high-energy electrons and positions.

Besides, one can also include the effect of substructures in the radio emission due to dark matter annihilation. The high density of dark matter subhaloes would enlarge the dark matter annihilation rate by a factor of $(1+B_{\rm boost})$, where $B_{\rm boost}$ is called the effective boost factor. The effective boost factor within $r=r_{\rm h}$ can be determined by \citep{Beck,Lavis}:
\begin{equation}
B_{\rm boost}(r_h) = 4\pi\int_0^{r_{\rm h}}r^2f_{\rm boost}\left(\frac{B(r)}{B_0}\right)\tilde{\rho}_{\rm sub}(r)dr
\end{equation}
where
\begin{equation}
\tilde{\rho}_{\rm sub}(r)=\frac{\rho_{\rm DM}(r)}{4\pi\int_0^{r_{\rm vir}}r^2\rho_{\rm DM}(r)dr}
\end{equation}
and $f_{\rm boost}$ is the total boost factor for the whole galaxy cluster. The total boost factor $f_{\rm boost}$ is redshift-dependent, which can be parameterised by \citep{Ando}:
\begin{eqnarray}
\log f_{\rm boost}& =&\frac{X(z)}{1+\exp[-a(z)(\log M_{\rm vir}-m_1(z))]}\nonumber\\
&&+c(z)\left(1+\frac{Y(z)}{1+\exp[-b(z)(\log M_{\rm vir}-m_2(z))]}\right)\nonumber\\
\end{eqnarray}
where for Okoli's concentration \citep{Okoli}
\begin{eqnarray}
X(z)&=&2.2e^{-0.75z}+0.67,\nonumber\\
Y(z)&=&2.5e^{-0.005z}+0.8,\nonumber\\
a(z)&=&0.1e^{-0.5z}+0.22,\nonumber\\
b(z)&=&0.8e^{-0.5(z-12)^4}-0.24,\nonumber\\
c(z)&=&-0.0005z^3-0.032z^2+0.28z-1.12,\nonumber\\
m_1&=&-2.6z+8.2,\nonumber\\
m_2&=&0.1e^{-3z}-12.
\end{eqnarray}
Here, $M_{\rm vir}=6.8\times 10^{14}M_{\odot}$ is the virial mass of RX J1720.1+2638 assuming the NFW density profile. For RX J1720.1+2638 cluster with redshift $z=0.164$ \citep{Biava}, the total boost factor is $f_{\rm boost}=18.82$. For the central small radio halo region with $r\le r_{\rm h}=70$ kpc, the effective boost factor is $B_{\rm boost}=1.31-1.44$ for $B_0=16.2-24.6$ $\mu$G. Since the value of the effective boost factor is small and there is no observational evidence showing the existence of substructures near the central small radio halo region, we do not consider the boost factor at first, but we will discuss the overall effect of the boost factor afterwards.

\section{Results and data analysis}
The radio spectral data of the central small radio halo in RX J1720.1+2638 cluster can be found in \citet{Giacintucci} (using GMRT and VLA) and \cite{Biava,Biava2} (using LOFAR). As the regions of interest for different observations are different, for a more consistent analysis, we mainly focus on the data obtained by GMRT and VLA (mainly covered the central small radio halo), which include the radio flux density with uncertainties $\sigma_i$ from frequency $\nu=0.317$ GHz to $\nu=8.44$ GHz (see Table 1). We will test several models and compare the predicted radio flux density $S(\nu)$ with the observed spectral data $S_i(\nu)$. The goodness of fit for different models can be determined by the $\chi^2$ value:
\begin{equation}
\chi^2=\sum_{i=1}^N \frac{[S(\nu)-S_i(\nu)]^2}{\sigma_i^2}.
\end{equation}
Since there are different number of free parameters involved in the models tested, we need to quantify the information loss due to increasing the number of parameters involved. We apply the Akaike information criterion (AIC) and Bayesian information criterion (BIC) to determine the best models in considering the goodness of fit with different number of free parameters. The corresponding criterion is defined as \citep{Davari}:
\begin{eqnarray}
{\rm AIC}&=&\chi^2+2k+\frac{2k(k+1)}{N-k-1},\\
{\rm BIC}&=&\chi^2+k\ln N,
\end{eqnarray}
where $N=6$ and $k$ are the number of data and the number of free parameters involved in the models respectively. The model with the smallest AIC and BIC values would be selected as the best models. Generally speaking, the AIC can help determine the best model among all tested models while the BIC can help determine the true model if the true model is included in the tested models.

Consider the radio emission originating only from the cosmic rays associated with the intracluster medium in RX J1720.1+2638. Generally speaking, there are several mechanisms to accelerate particles in the intracluster medium to produce synchrotron radio spectrum.  In particular, first-order Fermi acceleration (Fermi-I), second-order Fermi acceleration (Fermi-II), the primary emission models, and the secondary emission models are four popular emission mechanisms for galaxy clusters without Active Galactic Nuclei \citep{Rephaeli,Dennison,vanWeeren}. In view of these mechanisms, there are three functional forms which can best represent the radio spectra deriving from these different mechanisms, which are given by \citep{Jaffe,Rephaeli,Dennison,Schlickeiser,Thierbach}:
\begin{equation}
S_{\rm PL}(\nu)=S_0 \left(\frac{\nu}{\rm GHz} \right)^{-\alpha},
\end{equation}
\begin{equation}
S_{\rm Rep}(\nu)=S_0 \left(\frac{\nu}{\rm GHz} \right)^{\frac{2-\alpha}{2}} \left(\frac{1}{1+\nu/\nu_s} \right),
\end{equation}
\begin{equation}
S_{\rm insitu}(\nu)=S_0 \left(\frac{\nu}{\rm GHz} \right)^{\frac{3-\alpha}{2}} \exp \left(-\frac{\nu^{1/2}}{\nu_s^{1/2}} \right),
\end{equation}
where $S_0$, $\alpha$, and $\nu_s$ are empirical free parameters. Here, we call them `power-law form', `Rephaeli form', and `in-situ form' respectively according to the categorisation in \citet{Thierbach}. In fact, observations strongly suggest that turbulent acceleration can be found in some regions of RX J1720.1+2638 \citep{Mazzotta,Biava2}. The examples from magneto-hydrodynamical (MHD) turbulence can be categorised into Fermi-II process \citep{vanWeeren}. This is a stochastic process where particles scatter from magnetic inhomogeneities, which has been discussed in the studies of Coma cluster \citep{Schlickeiser,Thierbach}. The above three functional forms have been tested in the radio spectrum of Coma cluster, including the `power-law form' and `in-situ form' possibly associated with the Fermi-II process \citep{Jaffe,Thierbach}. Generally speaking, the above three functional forms can almost represent the possible shapes of the radio spectrum originated from cosmic rays \citep{Thierbach,vanWeeren}.

We fit the three benchmark cosmic-ray functional forms with the radio spectral data of RX J1720.1+2638. The corresponding AIC and BIC values can be found in Table 2. We can see that the power-law form has the smallest AIC and BIC values, which represents the best one among the three benchmark functional forms. Comparing with the power-law form, the difference in the AIC values from the other two forms is larger than 8.8. This means that the probability for these two forms being the best functional form is less than 2\% compared with the power-law form. Therefore, we select the power-law form as our reference functional form of the cosmic-ray model (CR-only model) for comparison with the addition of dark matter contribution. In Fig.~4, we show the spectral fits using different forms of the cosmic-ray model.

Next, we consider the dark matter-only model (DM-only model), which means that all the radio halo emission originates from dark matter annihilation. By setting $S(\nu)=S_{\rm DM}(\nu)$ and testing four major annihilation channels ($e$, $\mu$, $\tau$ and $b$), we can get the smallest AIC and BIC values for each annihilation channel. In fact, the effects of the parameters $\eta$ and $B_0$ within the possible ranges on $\chi^2$ are very small. Therefore, we particularly fix two sets of these parameters using their extreme values of the possible ranges, ($\eta=0.5$, $B_0=24.6$ $\mu$G) and ($\eta=1.0$, $B_0=16.2$ $\mu$G), to get the best scenarios. Here, we will test the DM-only model by using the NFW profile, Moore profile, and Einasto profile. The only free parameters in the fitting are the dark matter mass $m$ and the annihilation cross section $\langle \sigma v \rangle$ (i.e. $k=2$).

We find that the goodness of fits are almost the same for using the NFW profile, Moore profile, and Einasto profile. In Fig.~5, using the two fixed sets of $\eta$ and $B_0$, we plot $\chi^2$ against $m$ for four different popular annihilation channels (assuming the NFW profile). We can see that there exist some minimum $\chi^2$ values at particular $m$. This means that some best scenarios can be obtained for these annihilation channels. In Table 3, we summarise the AIC and BIC values for the best scenarios. In particular, $\tau$ channel with $m=15$ GeV and $b$ channel with $m=110$ GeV can give the smallest AIC and BIC values (following the NFW profile). Comparing the best scenario (i.e $b$ channel with $m=110$ GeV) with the CR-only model, the difference in AIC is $\Delta {\rm AIC} \approx 2.04$, which means that the CR-only model has only 36\% as probable as the best DM-only model to be the best model. This gives a considerable credence on showing that the DM-only model is the best model to account for the radio spectral data of the central radio halo of RX J1720.1+2638. Also, for considering BIC, we get $\Delta {\rm BIC} \approx 2.04 \ge 2$, which shows a marginally positive evidence for it being the true model. Moreover, our results reveal that the other two functional forms of cosmic rays (i.e. `Rephaeli form' and `in-situ form') are even more strongly rejected ($\Delta {\rm AIC}>10$). We show the corresponding spectral fits for the best scenarios in Fig.~6.

Finally, we consider the combined model, which represents the radio contribution of both dark matter annihilation and cosmic rays for the central radio halo of RX J1720.1+2638. We take the power-law form as the representative cosmic-ray model so that we have $S(\nu)=S_{\rm DM}(\nu)+S_{\rm PL}(\nu)$. This scenario involves both dark matter and cosmic-ray contributions, which is the most general scenario accounting for the radio halo spectrum. We first examine the scenario with the thermal annihilation cross section $\langle \sigma v \rangle=2.2 \times 10^{-26}$ cm$^3$/s \citep{Steigman} so that there are three free parameters involved ($m$, $S_0$ and $\alpha$). The plot $\chi^2$ against $m$ shown in Fig.~5 does not show any significant local minimum $\chi^2$, which means that there is no outstanding good fit for this specific scenario.

Then we test the combined model by taking $\langle \sigma v \rangle$ as a free parameter (i.e. $k=4$). Similarly, the goodness of fits are almost the same for using three different dark matter density profiles. Although adding one more free parameter can give some smaller $\chi^2$ values for a few annihilation channels (see Fig.~5), it also gives relatively large AIC values (see Table 4). Nevertheless, there exist a few cases in which the BIC values are smaller than the CR-only model (e.g. $e$ and $\mu$ channels). For the best scenario (i.e. $e$ channel with $m=10$ GeV following the Einasto profile), the difference in BIC compared with the CR-only model is $\Delta{\rm BIC}=6.1 \ge 2$, which represents a positive evidence for it being the true model. This suggests that adding dark matter contribution can slightly improve the entire fit of the radio spectral data (see Fig.~7). However, its large AIC value represents an extremely small probability for it being the best model. Moreover, as the data sample size is quite small ($N=6$), the formula for BIC in Eq.~(20) may not be a good approximation while the formula for AIC we used in Eq.~(19) can be applied for small data sample size. Therefore, the results drawn from AIC comparison are more reliable in this study. As a result, considering all of the above factors, the DM-only model is likely to be the best model to account for the radio halo spectrum, despite the fact that the CR-only model can also provide a good explanation for the observed radio spectrum.

Besides, we have also tested the effect of boost factor in our analysis. The overall effect of the boost factor is nearly negligible in the goodness of fits. It only decreases the best-fit values of annihilation cross section. 

\section{Discussion}
In this article, we have investigated on different models which can best-fit with the spectral data of the central radio halo in RX J1720.1+2638 cluster. The power-law cosmic-ray model can generally give a good fit with the radio data. However, the DM-only model with $b$ annihilation channel ($m=110$ GeV) and $\tau$ annihilation channel ($m=15$ GeV) can give slightly better fits for the radio data. These two scenarios can give smaller values of AIC and BIC which demonstrate considerable credence in accepting them to be the best models. The two best scenarios basically concur with some earlier best-fit results following the radio analysis of Large Magellanic Cloud (best-fit $m=50-280$ GeV via $b$ channel) \citep{Chan} and the analysis of gamma-ray excess at the Galactic Centre (best-fit $m \sim 10$ GeV via $\tau$ channel) \citep{Abazajian}. Besides, although the combined model with $e$ annihilation channel ($m=10$ GeV) can give the smallest BIC value among all models, it has a relatively large AIC value. As the radio data sample size is small, the AIC value would give a more reliable comparison for model selection. Therefore, our results suggest that the DM-only model is the best model to account for the central radio emission. If this is the case, dark matter contribution would be the dominant component in the radio emission at the centre of RX J1720.1+2638. Nevertheless, note that the functional forms tested for the cosmic-ray emission may not be exhaustive. It is possible that the emission due to turbulent acceleration slightly deviates from the three tested functional forms. In this case, cosmic-ray emission may be the best explanation for the observed radio spectrum for RX J1720.1+2638. Therefore, our conclusion is subject to the possible systematic uncertainties of the tested functional forms for the cosmic-ray emission.

Since the magnetic field strength would decrease when $r$ is large, the dark matter contribution would be greatly suppressed in the outskirt region of a galaxy cluster. Therefore, the best region that we can detect dark matter annihilation would be the central region. In other words, the central radio haloes in galaxy clusters would be excellent targets for constraining dark matter, like previous studies using the central radio halos of the Ophiuchus cluster \citep{Chan4} and Coma cluster \citep{Beck} to constrain dark matter.

Note that our best-fit annihilation cross sections (see Tables 3 and 4) might be in tension with some of the constraints obtained in recent studies, such as the gamma-ray constraints in Milky Way dwarf spheroidal galaxies \citep{Ackermann}, the best-fit ranges from the gamma-ray excess in the Galactic Centre \citep{Daylan}, and even the radio constraints obtained recently \citep{Beck}. However, as the annihilation cross section could be velocity-dependent \citep{Kiriu}, the constraints on the annihilation cross section obtained from analysing different structures cannot be directly compared with each other because the dark matter velocity dispersion would be significantly different in different structures, even in different galaxy clusters. Therefore, to better constrain the properties of dark matter annihilation cross section, a more comprehensive analysis across different structures involving velocity-dependent parameters is definitely required.

\begin{center}
\begin{table}
\caption{Integrated radio flux density for the central radio halo of RX J1720.1+2638, extracted from Table 4 of \citet{Giacintucci}}
\begin{tabular}{ |c|c|c|}
 \hline\hline
$\nu ({\rm GHz})$  &$S_i(\nu)({\rm mJy})$  & $\sigma_i$(mJy) \\
\hline
0.317 & 286 &38  \\
0.617& 144 &11  \\
1.28&59  &3  \\
1.48&60  &5  \\
4.86&18.7  &1.3  \\
8.44&6.2  &0.6  \\
\hline\hline
\end{tabular}
\label{observationdata}
\end{table}
\end{center}

\begin{table}
\caption{The smallest $\chi^2$, AIC, BIC values for the three benchmark functional forms of the cosmic-ray emission.}
\begin{tabular}{ |l|l|l|l|}
\hline\hline
 & Power-law & Rephaeli & In-situ \\
\hline
$\chi^2$ & 22.20 & 21.00 & 28.79 \\
AIC & 30.20 & 39.00 & 46.79 \\
BIC & 25.78 & 26.37 & 34.17 \\
\hline\hline
\end{tabular}
\end{table}

\begin{table}
\caption{The best-fit $m$, $\langle \sigma v \rangle$, $\chi^2$, AIC, and BIC values for four popular annihilation channels of the DM-only model, following the NFW, Einasto, and Moore density profiles respectively. The parameter set of ($B_0=24.6$ $\mu$G, $\eta=0.5$) would give the smallest $\chi^2$ values.}
\begin{tabular}{ |l|l|l|l|l|}
\hline\hline
NFW & $e$ & $\mu$ & $\tau$ & $b$ \\
\hline
$m$ (GeV) & 5 & 10 & 15 & 110 \\
$\langle \sigma v \rangle$ ($10^{-26}$ cm$^3$/s) & 9.9 & 111 & 469 & 3882 \\
$\chi^2$ & 31.54 & 23.09 & 20.18 & 20.15 \\
AIC & 39.54 & 31.09 & 28.18 & 28.15 \\
BIC & 35.12 & 26.67 & 23.76 & 23.74 \\
\hline\hline
\end{tabular}
\begin{tabular}{ |l|l|l|l|l|}
\hline\hline
Einasto & $e$ & $\mu$ & $\tau$ & $b$ \\
\hline
$m$ (GeV) & 5 & 10 & 15 & 110 \\
$\langle \sigma v \rangle$ ($10^{-26}$ cm$^3$/s) & 25.08 & 284.5 & 1205 & 9977 \\
$\chi^2$ & 40.17 & 23.42 & 21.08 & 20.12 \\
AIC & 48.17 & 31.42 & 29.08 & 28.12 \\
BIC & 43.75 & 27.01 & 24.66 & 23.70 \\
\hline\hline
\end{tabular}
\begin{tabular}{ |l|l|l|l|l|}
\hline\hline
 Moore& $e$ & $\mu$ & $\tau$ & $b$ \\
\hline
$m$ (GeV) & 5 & 10 & 15 & 90 \\
$\langle \sigma v \rangle$ ($10^{-26}$ cm$^3$/s) & 0.462 & 5.39 & 22.33 & 155.08 \\
$\chi^2$ & 38.27 & 33.93 & 21.81 & 20.25 \\
AIC & 46.27 & 41.93 & 29.81 & 28.25 \\
BIC & 41.85 & 37.51 & 25.39 & 23.83 \\
\hline\hline
\end{tabular}
\end{table}

\begin{table}
\caption{The best-fit $m$, $\langle \sigma v \rangle$, $\chi^2$, AIC, and BIC values for four popular annihilation channels of the combined model, following the NFW, Einasto, and Moore density profiles respectively. The parameter set of ($B_0=16.2$ $\mu$G, $\eta=1$) would give the smallest $\chi^2$ values.}
\begin{tabular}{ |l|l|l|l|l|}
\hline\hline
NFW & $e$ & $\mu$ & $\tau$ & $b$ \\
\hline
$m$ (GeV) & 10 & 15 & 25 & 150 \\
$\langle \sigma v \rangle$ ($10^{-26}$ cm$^3$/s) & 17.8 & 199 & 1020 & 4204 \\
$\chi^2$ & 15.44 & 16.84 & 17.95 & 20.09 \\
AIC & 63.44 & 64.84 & 65.95 & 68.09 \\
BIC & 22.61 & 24.00 & 25.12 & 27.26 \\
\hline\hline
\end{tabular}
\begin{tabular}{ |l|l|l|l|l|}
\hline\hline
Einasto & $e$ & $\mu$ & $\tau$ & $b$ \\
\hline
$m$ (GeV) &10  &15  &20  &240  \\
$\langle \sigma v \rangle$ ($10^{-26}$ cm$^3$/s) & 61.82 &337.9  &83.82  &25130  \\
$\chi^2$ &12.50  &18.58  &21.24  &19.58  \\
AIC &60.50  &66.58  &69.24  &67.58  \\
BIC & 19.67 & 25.75 &28.40  &26.75  \\
\hline\hline
\end{tabular}
\begin{tabular}{ |l|l|l|l|l|}
\hline\hline
 Moore& $e$ & $\mu$ & $\tau$ & $b$ \\
\hline
$m$ (GeV) &5  &10  &20  &160  \\
$\langle \sigma v \rangle$ ($10^{-26}$ cm$^3$/s) & 0.154 &0.217  &25.3  &229.68  \\
$\chi^2$ & 17.99 &21.01  &18.34  &19.52  \\
AIC & 65.99 &69.01  &66.34  &67.52  \\
BIC &25.16  &28.18  &25.50  &26.68  \\
\hline\hline
\end{tabular}
\end{table}

\begin{figure}
\begin{center}
\vskip 5mm
\includegraphics[width=80mm]{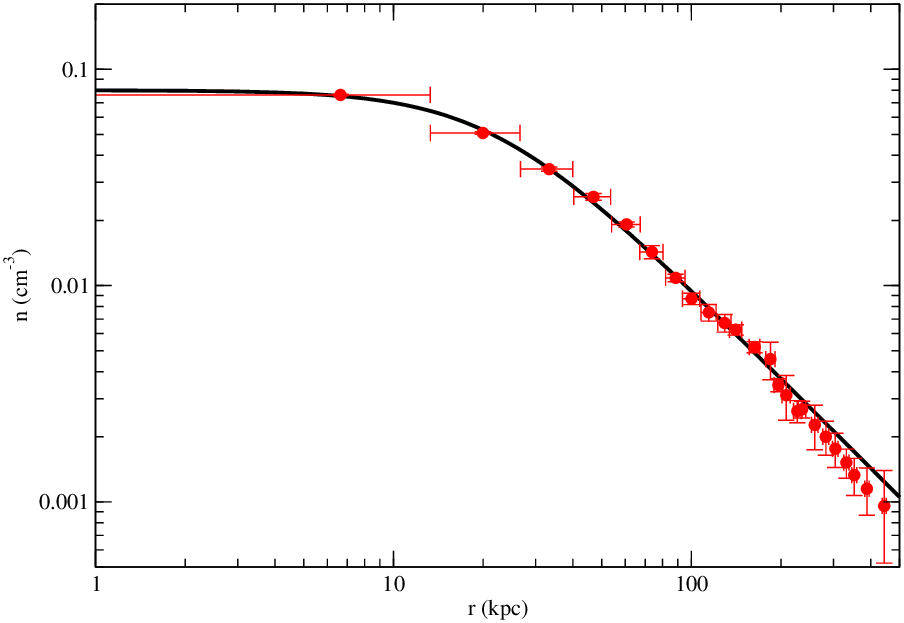}
\caption{The best-fit thermal electron density profile assuming the single-$\beta$ model. The best-fit parameters are $n_0=0.08$ cm$^{-3}$, $r_c=21.7$ kpc, and $\beta=0.46$. The data with $1\sigma$ uncertainties are extracted from \citet{Cavagnolo}.}
\label{Fig1}
\end{center}
\end{figure}

\begin{figure}
\begin{center}
\vskip 5mm
\includegraphics[width=80mm]{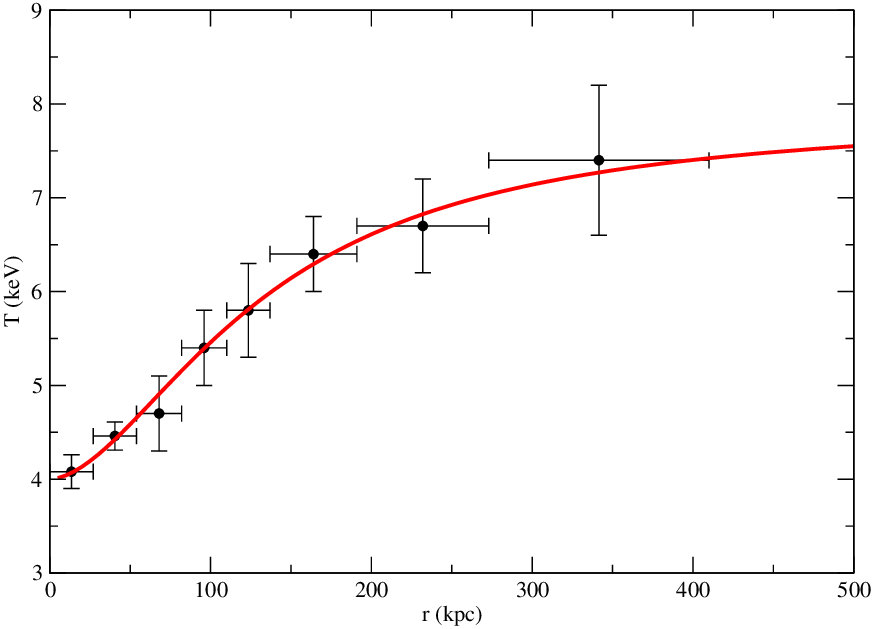}
\caption{The best-fit temperature profile of RX J1720.1+2638 cluster using Eq.~(8). The best-fit parameters are $T_0=4.0$ keV, $T_1=3.9$ keV, $r_0=134$ kpc, and $a=1.76$. The data with $1\sigma$ uncertainties are extracted from \citet{Jimenez}.}
\label{Fig2}
\end{center}
\end{figure}

\begin{figure}
\begin{center}
\vskip 5mm
\includegraphics[width=80mm]{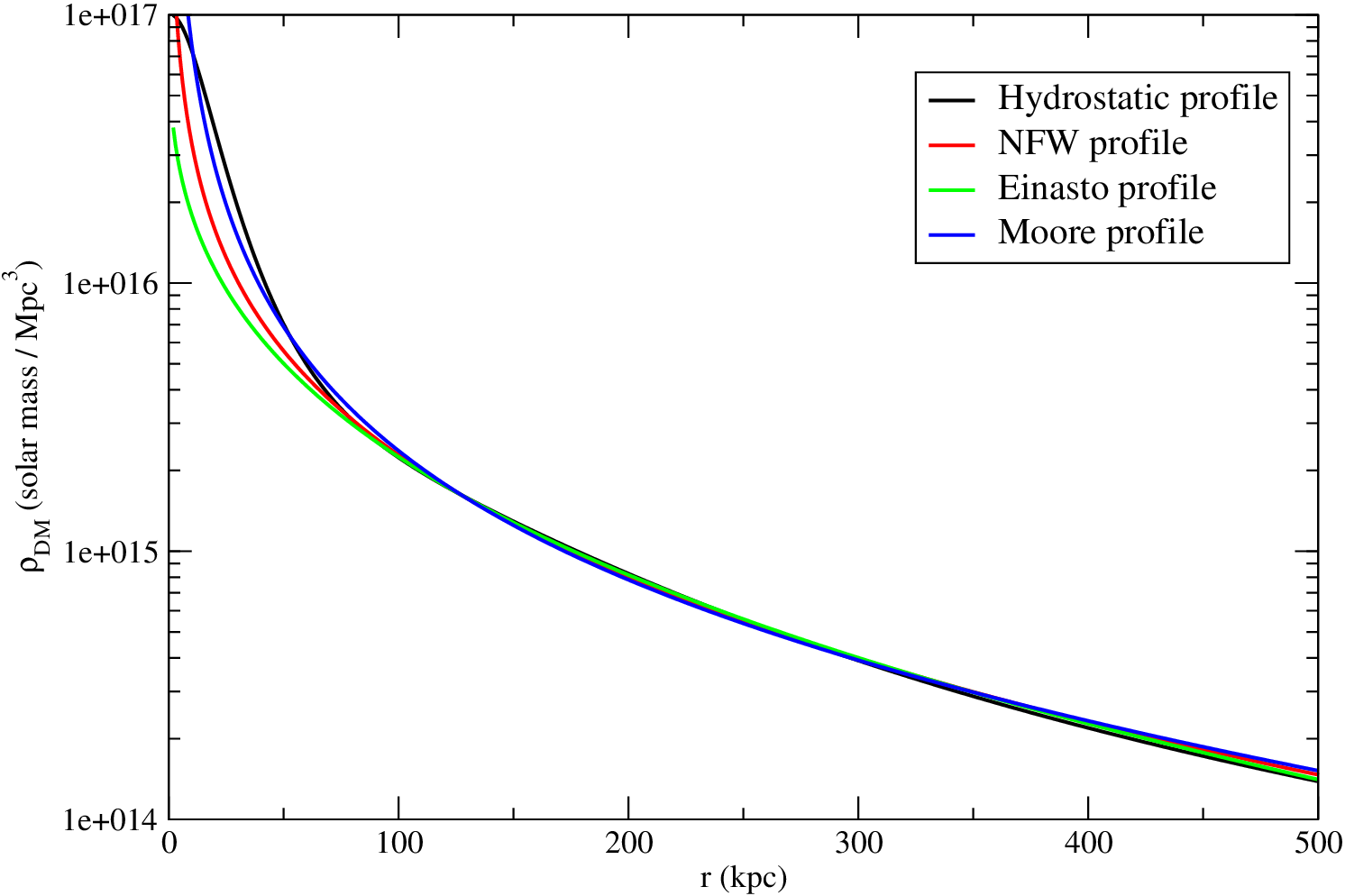}
\caption{The NFW profile (red), Einasto profile (green), Moore profile (blue), and hydrostatic profile (black) for the dark matter density in RX J1720.1+2638.}
\label{Fig3}
\end{center}
\end{figure}

\begin{figure}
\begin{center}
\vskip 5mm
\includegraphics[width=80mm]{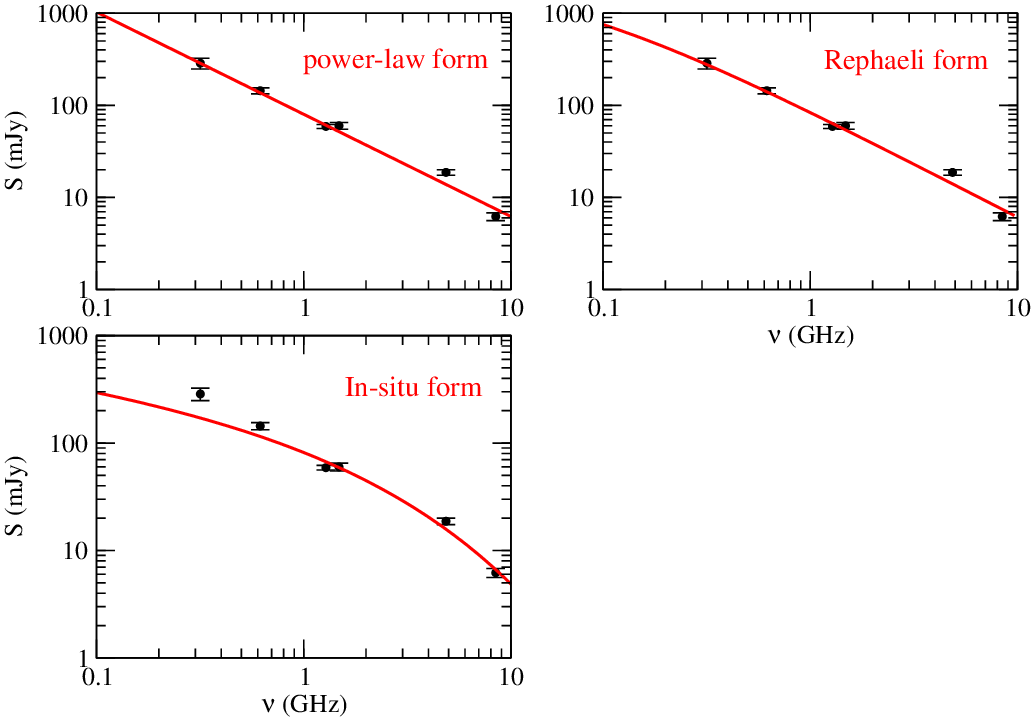}
\caption{The radio spectra for the best-fit scenarios of the CR-only model. Top left: the power-law form; Top right: the Rephaeli form; Bottom: the in-situ form. The data with $1\sigma$ uncertainties are extracted from \citet{Giacintucci}.}
\label{Fig4}
\end{center}
\end{figure}

\begin{figure}
\begin{center}
\vskip 5mm
\includegraphics[width=80mm]{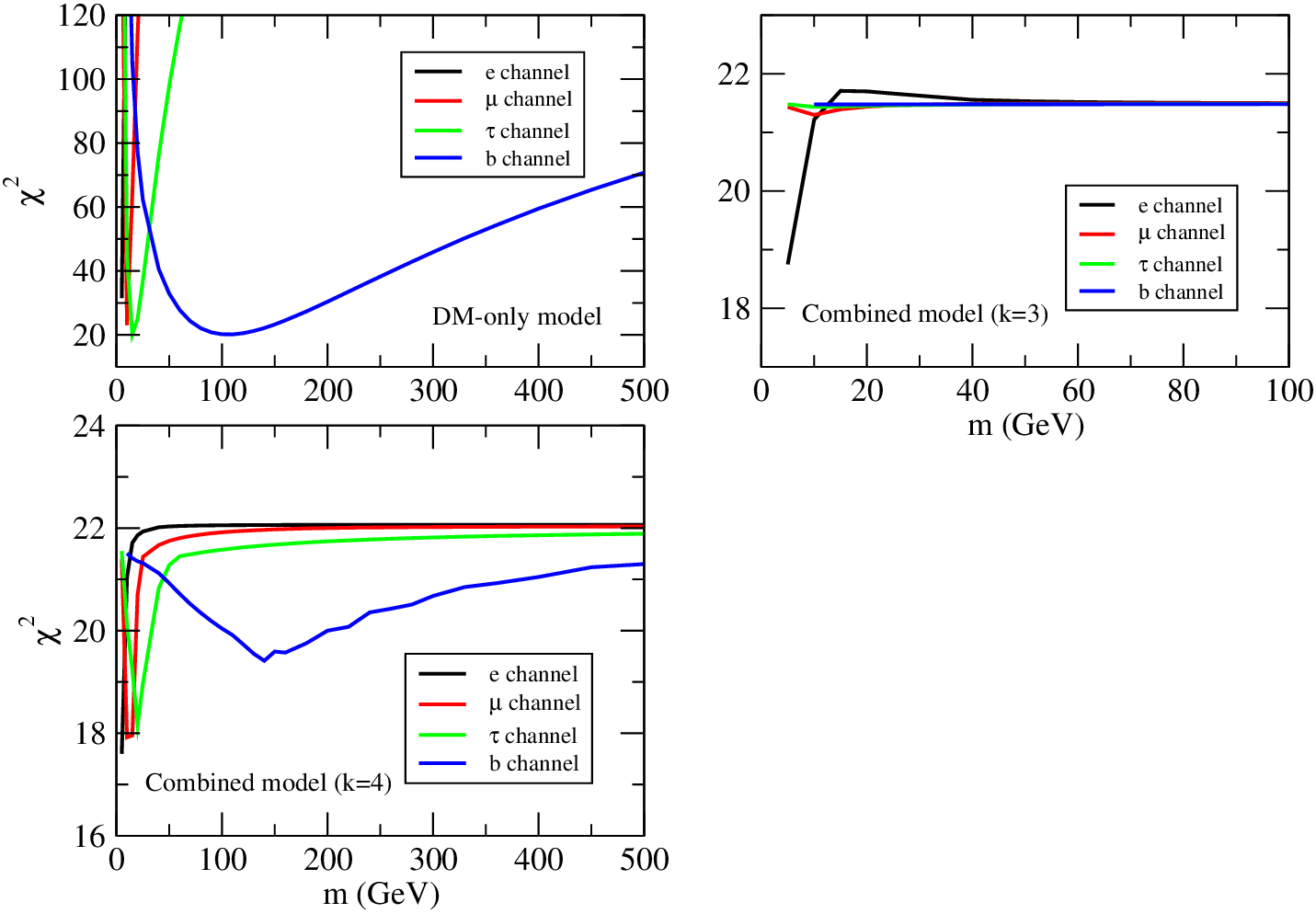}
\caption{$\chi^2$ against $m$ for four different annihilation channels (all following the NFW profile). Top left: the DM-only model; Top right: the combined model with the thermal annihilation cross section ($k=3$); Bottom: the combined model with the annihilation cross section being a free parameter ($k=4$).}
\label{Fig5}
\end{center}
\end{figure}

\begin{figure}
\begin{center}
\vskip 5mm
\includegraphics[width=80mm]{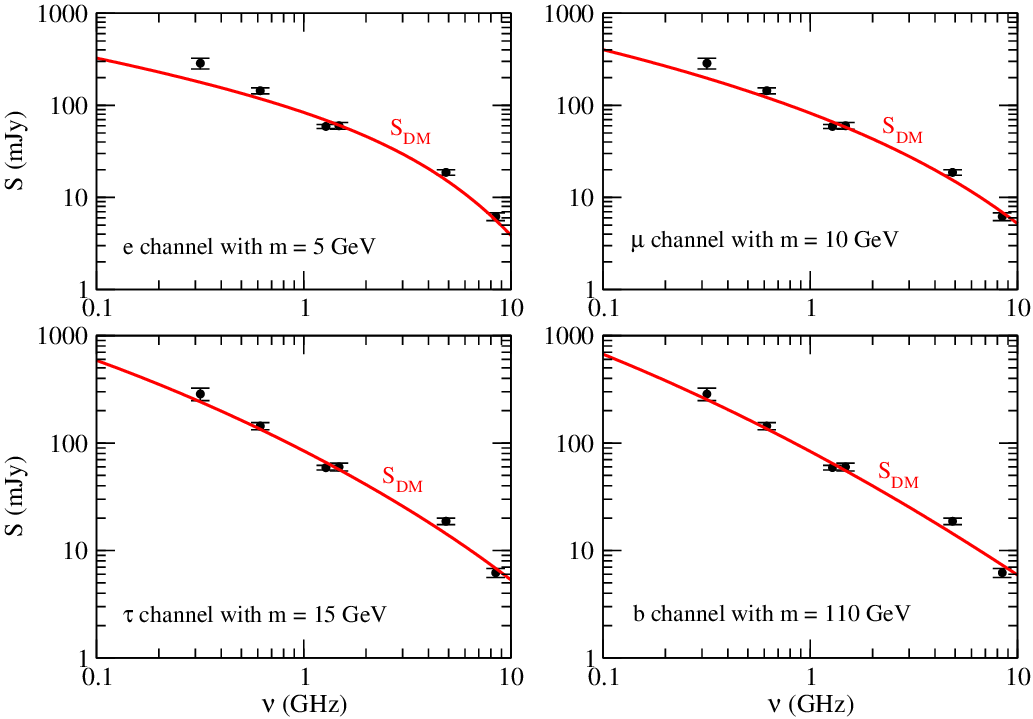}
\caption{The radio spectra for the best-fit scenarios of the DM-only model (all following the NFW profile and $B_0=24.6$ $\mu$G). Top left: $e$ channel with $m=5$ GeV; Top right: $\mu$ channel with $m=10$ GeV; Bottom left: $\tau$ channel with $m=15$ GeV; Bottom right: $b$ channel with $m=110$ GeV. The data with $1\sigma$ uncertainties are extracted from \citet{Giacintucci}.}
\label{Fig6}
\end{center}
\end{figure}

\begin{figure}
\begin{center}
\vskip 5mm
\includegraphics[width=80mm]{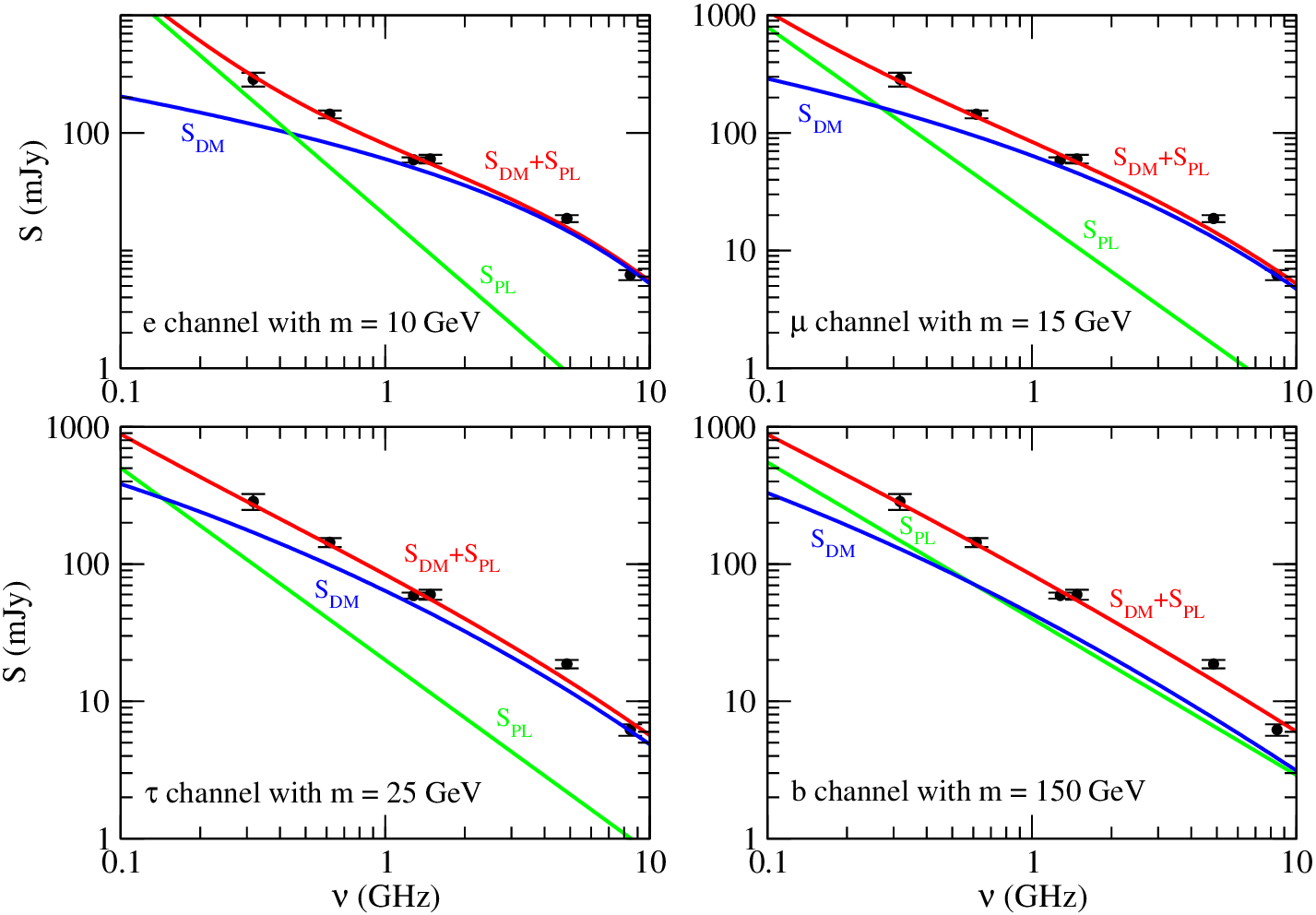}
\caption{The radio spectra for the best-fit scenarios of the combined model ($B_0=16.2$ $\mu$G). The blue, green, red lines represent the dark matter contribution, cosmic-ray contribution (following the power-law form), and the total combined radio flux density respectively. Top left: $e$ channel with $m=10$ GeV (Einasto profile); Top right: $\mu$ channel with $m=15$ GeV (NFW profile); Bottom left: $\tau$ channel with $m=25$ GeV (NFW profile); Bottom right: $b$ channel with $m=150$ GeV (NFW profile). The data with $1\sigma$ uncertainties are extracted from \citet{Giacintucci}.}
\label{Fig7}
\end{center}
\end{figure}

\section{Acknowledgements}
We thank the anonymous referees for useful comments. The work described in this paper was partially supported by the Dean's Research Fund (0401E) and a grant (RG 3/2023-2024R) from the Education University of Hong Kong and the grant from the Research Grants Council of the Hong Kong Special Administrative Region, China (Project No. EdUHK 18300922).

\section{Data availability}
The data underlying this article will be shared on reasonable request to the corresponding author.


\label{lastpage}

\end{document}